\documentstyle[11pt]{article}
\input{epsf}
\def\reference{\parskip 0pt\par\noindent\hangindent 0.5 truecm}

\def\arcsec{\hbox{$^{\prime\prime}$}}
\def\etal{et al.\ }
\def\ie{i.e.\ }
\def\mgii{Mg\,{\sc ii}}
\begin{document}

%
% Title
% Capitalise the title normally - do not use ALL CAPS.
%
\title{The Optical Emission from Gamma-Ray Quasars}
%

% Authors
% Here comes the author(s) of the paper. Please add the appropriate author
% names for your paper and indicate within the $^...$ the number(s)
% which corresponds to the institute(s) of each author. In this example
% the second author has two institutional affiliations.
% Add or remove authors as required.
% **** IMPORTANT: Leave the closing curly bracket line as is. ******

\author{M.T. Whiting$^{1}$,
 P. Majewski$^{2}$ and
 R.L. Webster$^{2}$
} % IMPORTANT: leave this curly bracket as the first character of this line.

% Date - leave this blank.
\date{}
\maketitle

% Institutions
% Here fill in your institute name(s) and address(es)
% The number in $^...$ indicates the author number.  For example
{\center $^1$ Department of Astrophysics and Optics, School of Physics,
  University of New South Wales, Sydney NSW 2052, Australia
  \\ mwhiting@phys.unsw.edu.au \\[3mm]
$^2$ Astrophysics Group, School of Physics, University of Melbourne,
  Victoria 3010, Australia\\rwebster@physics.unimelb.edu.au\\[3mm]
Received 2001 November 29, accepted 2003 April 23}
% Abstract
% Simply place your abstract between the \begin{abstract} and
% \end{abstract} commands.
%
\begin{abstract}
We present photometric observations of six radio-loud quasars that were detected by the COMPTEL gamma-ray
telescope. The data encompass seven wavebands in the optical and near infrared. After correction for Galactic
extinction, we find a wide range in optical slopes. Two sources are as blue as optically-selected quasars, and are
likely to be dominated by the accretion disc emission, while three others show colours consistent with a red
synchrotron component. We discuss the properties of the COMPTEL sample of quasars, as well as the implications our
observations have for multiwavelength modelling of gamma-ray quasars.
\end{abstract}

{\bf Keywords:} galaxies: active --- quasars: individual (PKS\,0208$-$512, PKS\,0506$-$612, PKS\,0528$+$134,
PKS\,1622$-$297, PKS\,2230$+$114, PKS\,2251$+$158)
% Place keywords here. Please write all keywords in lower case. PASA uses the
 %standard list of subject
% headings adopted by The Astrophysical Journal and available from URL:
%   http://www.journals.uchicago.edu/ApJ/keywords_text.html

% A formatting command to add space between the author list and the body
% of the paper when printed. This spacing may be changed as desired.
\bigskip

%
% Body of paper
%

\section{Introduction}

Active galactic nuclei (AGN) are powerful emitters in all wavebands. Emission processes from different parts of
the spectrum in many cases appear to be linked. For example, the electrons in the jet which produce the
synchrotron radiation may at the same time upscatter photons into the X-ray and gamma-ray regime via the inverse
Compton (IC) process (see for example the analysis by Bloom \& Marscher 1996). In order to unravel the
contribution from each emission process, the shape of each distinct emission component needs to be determined, and
the associated physical parameters measured.

The IC emission is produced by seed photons (internal, from the jet itself, or external, from the accretion disc
or the emission-line clouds) which are upscattered by the electrons (or other relativistic particles) in the jet.
The energy of the photons increases by a factor of $\sim \gamma^2$, where $\gamma$ is the Lorentz factor of the
relativistic particle. Consequently, there are two crucial factors that determine the nature of the IC emission:
the maximum energy of the jet electrons, which determines the turnover frequency of the synchrotron component; and
the typical energy of the seed photons, which is governed by the relative strengths of the thermal and non-thermal
components. The optical regime is where these components are both energetically important, and so a good
understanding of the emission at optical frequencies is crucial for accurate modelling of the high-energy
emission.

Before COMPTEL was shut down in June 2000, this telescope detected nine blazars in the range 0.75 -- 30 MeV which
were also detected by EGRET in the gamma-ray regime $\ge$ 100 MeV (Sch{\"o}nfelder \etal 2000; Collmar 2001). For
many of the sources in this sample, the data coverage in the optical/near infrared (NIR) region is very poor. To
better understand this important part of the spectral energy distribution (SED), six blazars from the COMPTEL
sample, chosen subject to visibility during an observation period from 27 August to 2 September 2001, were
observed in the photometric bands $B$, $V$, $R$, $I$, $J$, $H$, and $K_n$. Basic information for this sample is
shown in Table~\ref{tab-sample}. We note that the redshift for PKS\,1622$-$297 ($z=0.815$) is taken from the
Parkes Catalogue (Wright \& Otrupcek 1990, hereafter PKSCAT90), wherein no reference is given. No other
determination of its redshift has been made in the literature, and, while we use this value herein, it should be
regarded as potentially suspect.

\begin{table}
\caption{Sample description. E$(B-V)$ values are the Galactic
  extinction values from Schlegel et al.\ (1998).}
\begin{tabular}{cccccc}
\hline
Name           & Other    & RA          & DEC            &E$(B-V)$ & z  \\
               & name     &(J2000)      & (J2000)        & (mag)   &    \\
\hline \hline
PKS\,0208$-$512 & ---      & 02:10:46.20 & $-$51:01:01.9  &0.020    & 0.999 \\
PKS\,0506$-$612 & ---      & 05:06:44.00 & $-$61:09:41.0  &0.025    & 1.093 \\
PKS\,0528$+$134 & ---      & 05:30:56.36 & $+$13:31:55.0  &0.860    & 2.060  \\
PKS\,1622$-$297 & ---      & 16:26:06.02 & $-$29:51:27.0  &0.454    & 0.815 \\
PKS\,2230$+$114 & CTA 102  & 22:32:36.40 & $+$11:43:51.3  &0.072    & 1.037 \\
PKS\,2251$+$158 & 3C 454.3 & 22:53:57.75 & $+$16:08:53.6  &0.105    & 0.859 \\
\hline
\end{tabular}
\label{tab-sample}
\end{table}

Detailed models of the multiwavelength spectra of several of the sources in this sample have been published in the
literature.  For example, Mukherjee \etal (1999) describe a model with both synchrotron and IC emission components
fitted to nine epochs of monitoring data for the source PKS\,0528$+$134.  While considerable effort has been
expended in fitting the high-energy spectrum (X-rays and above) of this source, the information in the low-energy
part of the spectrum, from optical to radio energies, has largely been ignored. Rather than modelling one source
in detail, by considering a modest sample of six sources more general conclusions can be drawn about the nature of
the emission processes.

As synchrotron emission causes linear polarisation, optical
polarisation measurements can be used to check the predictions from
the modelling. Optical polarimetry of most of the sources in this
sample is available from the literature. It will be shown that the
sources with high optical polarisation are generally distinguished by
their radio to optical spectral shapes, and are consistent with having
a synchrotron component extending to optical wavelengths.

Section \ref{sec-data} describes the photometric optical and NIR data for the six sources.  The SEDs in each case
are adequately modelled by a power law in Section \ref{sec-model}. Finally, in Section \ref{disc} we examine the
implications of our observations for multiwavelength modelling of gamma-ray sources, as well as the demographics
of the sample as a whole.

\section{Observations and Data}
\label{sec-data}

\subsection{Observations}

The six blazars were observed in optical bands $B$, $V$, $R$, and $I$ on the night of 29 August 2001 with the ANU
40\,in telescope at Siding Spring Observatory (SSO). The exposure time in each band was 300 seconds. The seeing
was typically $2.5\arcsec - 3\arcsec$, but conditions were generally photometric (although the weather did force
some breaks in the observing, and the loss of the $B$ band data for PKS\,1622$-$297). Cross-checking of field
stars from Raiteri \etal (1998) enables verification of the photometric conditions.  Observations of standard
stars from Graham (1982) enabled calibration to an absolute magnitude/flux scale.

Observations in the NIR bands $J$, $H$, and $K_n$ were made on the nights of 31 August and 1 September 2001 with
the CASPIR (Cryogenic Array Spectrometer/Imager---see McGregor \etal 1994) instrument on the ANU 2.3\,m telescope,
SSO. Each image is a median-combination of four (or, in the case of the fainter sources, 20) dithered images of 60
seconds exposure time each (these individual exposures were: 2 cycles of 30 seconds for $J$; 6 cycles of 10
seconds in $H$; and 12 cycles of 5 seconds in $K_n$).

Poor weather on the first night allowed us to observe only PKS\,1622$-$297 and two standard stars, but the second
night had good conditions, allowing us to observe the remaining sources. The seeing was better than for the
optical (typically $\approx 1.5\arcsec-2\arcsec$) and conditions were generally photometric.  Individual frames
where the background increased due to the presence of cloud were discarded prior to co-addition. Observations of
IRIS standard stars from the Carter system (Carter \& Meadows 1995) were used for calibration.

\subsection{Data}

The data reduction was performed in a standard way using IRAF
software. Optical images were bias- and overscan-subtracted, and then
flat-fielded. The flat-field images were generated from a combination
of night sky flats (created from all object images such that any
sources are excluded) and flats made from exposures of a uniformly
cloudy sky.  NIR images were bias- and dark-subtracted, and then
flat-fielded using flats derived from the difference of exposures of
the dome with and without illumination. This removes telescope
emission and improves photometric accuracy. The sky emission was then
subtracted from the individual images using a median of all dither
positions (excluding those with high background due to cloud), and
then the individual dither positions were added together to produce
the final image.

The photometry was performed using the IRAF {\it daophot} package. The resulting optical and NIR magnitudes are
shown in Tables \ref{tab-opt} and \ref{tab-nir} respectively. The errors quoted are $1\sigma$ errors resulting
from the sum in quadrature of the random photometric errors, as calculated by the {\it daophot} package, and the
systematic errors that result from the spread in zero points for the various standard star observations over the
course of the night. The photometry includes corrections for the varying atmospheric extinction due to the
different airmasses of the sources. For two sources (PKS\,0506$-$612 and PKS\,1622$-$297), subtraction of nearby
stars was required to provide accurate background subtraction for the aperture photometry (although in no case
were the images of the star and quasar overlapping).  Figure \ref{fig-fits} plots the observed photometric SED for
each source as a function of observed wavelength.

The magnitudes were converted into fluxes to enable modelling and plotting of the data. The zero magnitude fluxes
for the optical bands were taken from Bessell, Castelli, \& Plez (1998), while the CASPIR zero magnitudes are
obtained from a 11200\,K blackbody normalised to $F_\lambda(555 {\rm nm})=3.44\times
10^{-12}$\,W\,cm$^{-2}$\,$\mu{\rm m}^{-1}$ (Bersanelli, Bouchet, \& Falomo 1991). The values of the zero points
are shown in Table \ref{tab-zeropt}. Note that although the observations used the $K_n$ filter, they were
calibrated according to the $K$ magnitudes of the IRIS standards without a colour-correction term, and so have
been normalised using the $K$ band zero point.

\subsection{Galactic Extinction}

Values for the Galactic extinction, taken from the maps of Schlegel, Finkbeiner, \& Davis (1998) are included in
Table \ref{tab-sample}. For two objects, PKS\,0528$+$134 and PKS\,1622$-$297, the extinction is large due to their
location close to the Galactic plane. We note that Zhang \etal (1994) found evidence for excess absorption over
the Galactic value for PKS\,0528$+$134 from X-ray observations, which would imply a larger $E(B-V)$ value than
that from Schlegel \etal (1998) (and hence a larger correction to the observed magnitudes). For our analysis,
however, we just use the Galactic value.

The values of E$(B-V)$ were transformed to an absorption $A_\lambda$ for each photometric band using the
conversion factors given in Schlegel \etal (1998). These absorption values were then subtracted from the observed
magnitudes to give the extinction-corrected magnitudes, and both sets of data (raw and corrected) are included in
Tables \ref{tab-opt} and \ref{tab-nir}.  The photometric extinction-corrected SEDs for each object have also been
plotted in Figure \ref{fig-fits}.

\begin{table}
\caption{Optical magnitudes and dates of observation. For each source, the first row shows the observed
magnitudes, while the second row shows the magnitudes after correction for Galactic extinction (see text).}
\begin{tabular}{lccccc}
\hline
Source & $B$ & $V$ & $R$ & $I$ &UT date  \\
\hline PKS\,0208$-$512 & 16.27$\pm$0.04 & 15.80$\pm$0.03 & 15.43$\pm$0.03 &
14.94$\pm$0.03 & 29 Aug 2001  \\
               & 16.18$\pm$0.04 & 15.73$\pm$0.03 & 15.38$\pm$0.03 &
14.90$\pm$0.03 &  \\ \\
PKS\,0506$-$612 & 17.67$\pm$0.07 & 17.25$\pm$0.05 & 16.88$\pm$0.04 &
16.62$\pm$0.05 & 29 Aug 2001  \\
               & 17.56$\pm$0.07 & 17.17$\pm$0.05 & 16.82$\pm$0.04 &
16.57$\pm$0.05 &  \\ \\
PKS\,0528$+$134 & 20.76$\pm$0.23 & 19.53$\pm$0.15 & 18.98$\pm$0.11 &
17.94$\pm$0.11 & 29 Aug 2001  \\
               & 17.04$\pm$0.23 & 16.74$\pm$0.15 & 16.72$\pm$0.11 &
16.25$\pm$0.11 & \\ \\
PKS\,1622$-$297 & ---            & 18.84$\pm$0.33 & 18.29$\pm$0.25 &
17.88$\pm$0.24 & 29 Aug 2001  \\
               & ---            & 17.37$\pm$0.33 & 17.09$\pm$0.25 &
16.99$\pm$0.24 &  \\  \\
PKS\,2230$+$114 & 17.20$\pm$0.05 & 16.82$\pm$0.03 & 16.43$\pm$0.04 &
16.03$\pm$0.04 & 29 Aug 2001  \\
               & 16.89$\pm$0.05 & 16.59$\pm$0.03 & 16.24$\pm$0.04 &
15.89$\pm$0.04 & \\  \\
PKS\,2251$+$158 & 15.34$\pm$0.05 & 14.83$\pm$0.01 & 14.43$\pm$0.03 &
13.88$\pm$0.03 & 29 Aug 2001  \\
               & 14.88$\pm$0.05 & 14.49$\pm$0.01 & 14.15$\pm$0.03 &
13.67$\pm$0.03 & \\  \\
\hline
\end{tabular}
\label{tab-opt}
\end{table}

\begin{table}
\caption{NIR magnitudes and dates of observation. For each source, the first row shows the observed magnitudes,
while the second row shows the magnitudes after correction for Galactic extinction (see text).}
\begin{tabular}{lcccc}
\hline
Source & $J$ & $H$ & $K$ & UT date  \\
\hline PKS\,0208$-$512 & 13.77$\pm$0.03 & 12.93$\pm$0.01 & 12.18$\pm$0.02 & 1
Sept 2001  \\
               & 13.75$\pm$0.03 & 12.92$\pm$0.01 & 12.17$\pm$0.02 & \\  \\
PKS\,0506$-$612 & 16.68$\pm$0.04 & 16.07$\pm$0.03 & 15.16$\pm$0.05 & 1
Sept 2001  \\
               & 16.66$\pm$0.04 & 16.06$\pm$0.03 & 15.15$\pm$0.05 & \\  \\
PKS\,0528$+$134 & 16.99$\pm$0.09 & 16.16$\pm$0.06 & 15.26$\pm$0.07 & 1
Sept 2001 \\
               & 16.21$\pm$0.09 & 15.66$\pm$0.06 & 14.94$\pm$0.07 & \\ \\
PKS\,1622$-$297 & 16.45$\pm$0.03 & 15.94$\pm$0.03 & 15.31$\pm$0.03 & 31
Aug 2001   \\
               & 16.04$\pm$0.03 & 15.68$\pm$0.03 & 15.14$\pm$0.03 & \\  \\
PKS\,2230$+$114 & 15.25$\pm$0.05 & 14.48$\pm$0.04 & 13.73$\pm$0.06 & 1
Sept 2001  \\
               & 15.18$\pm$0.05 & 14.44$\pm$0.04 & 13.70$\pm$0.06 & \\  \\
PKS\,2251$+$158 & 13.06$\pm$0.03 & 12.14$\pm$0.01 & 11.31$\pm$0.03 & 1
Sept 2001  \\
               & 12.96$\pm$0.03 & 12.08$\pm$0.01 & 11.27$\pm$0.03 & \\  \\
\hline
\end{tabular}
\label{tab-nir}
\end{table}

\begin{table}
\caption{Filters used, with their central wavelengths and the zero magnitude fluxes used for flux conversion.}
\begin{tabular}{lcc}
\hline
Filter &Flux of zero magnitude    &$\lambda$\\
       &star ($F_\lambda$, W cm$^{-2}$ $\mu{\rm m}^{-1}$)   &$(\mu{\rm m})$\\
\hline
$B$    &$6.32\times 10^{-8}$  &0.44\\
$V$    &$3.63\times 10^{-8}$  &0.55\\
$R$    &$2.18\times 10^{-8}$  &0.70\\
$I$    &$1.13\times 10^{-8}$  &0.88\\
$J$    &$3.11\times 10^{-9}$  &1.239\\
$H$    &$1.15\times 10^{-9}$  &1.649\\
$K$    &$4.10\times 10^{-10}$ &2.192\\
\hline
\end{tabular}
\label{tab-zeropt}
\end{table}

\section{Model Fitting}
\label{sec-model}

\begin{figure}[h]
\epsfbox{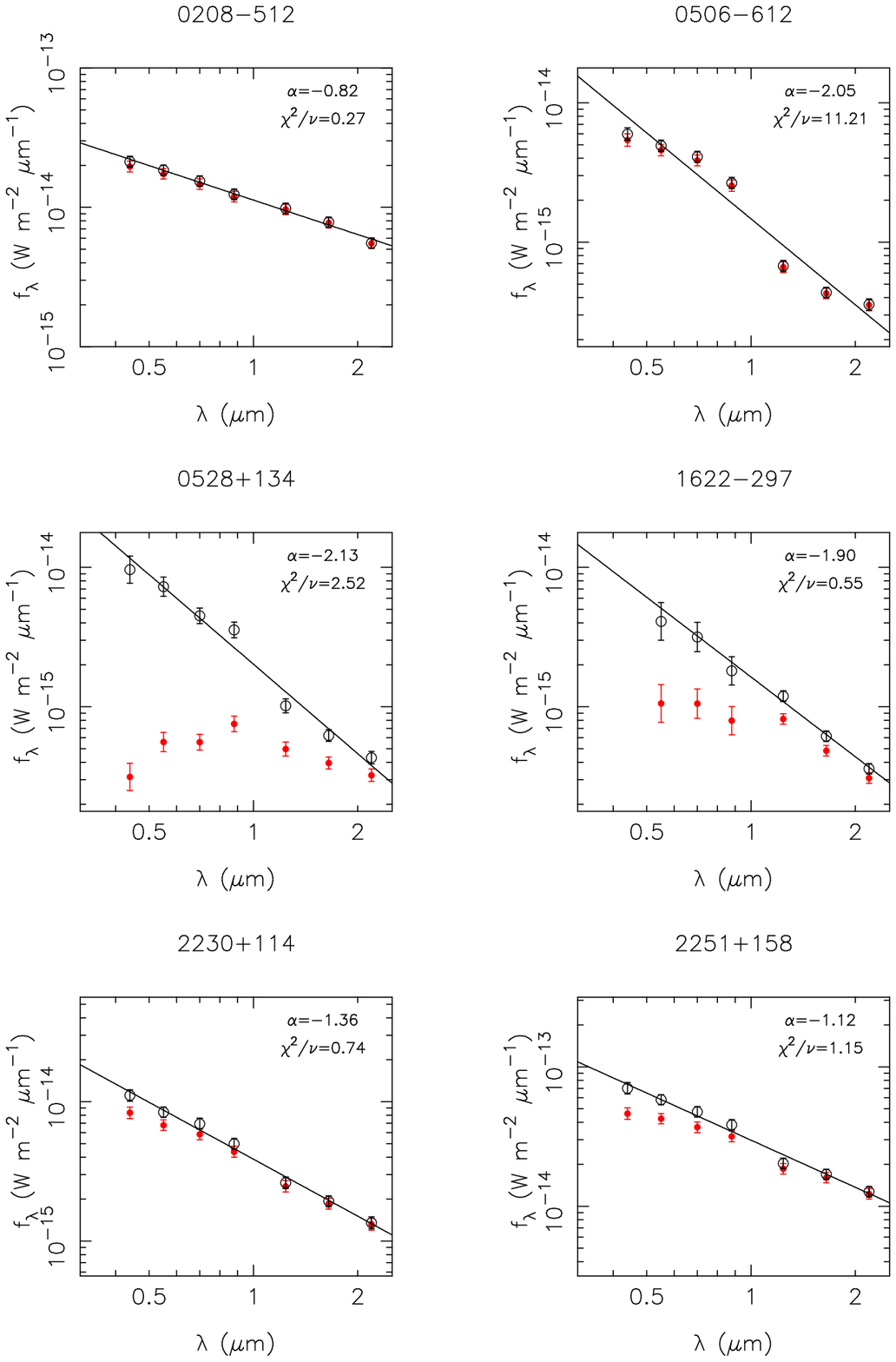}
\caption{Photometry and model fits for sources. The observed
  photometry is shown by the red solid symbols, whereas the photometry
  corrected for Galactic extinction is shown with the open
  symbols. The error bars shown are $1\sigma$. The fit to the
  corrected photometry is shown by the line, with the parameters of
  the fit (power law index and reduced-$\chi^2$) indicated in the
  corner of each plot. Note that each plot has a vertical scale
  spanning two decades, although the range is different for each
  source.}
\label{fig-fits}
\end{figure}

Each of the extinction-corrected SEDs has been fitted by a power law, given by $f_\lambda(\lambda) \propto
\lambda^\alpha$, using a $\chi^2$-minimisation method. The fit is deemed `acceptable' if the reduced $\chi^2$
(i.e.\ $\chi^2/\nu$, where $\nu$ is the number of degrees of freedom in the fitted model) is below the 99\%
confidence level. For the power-law model fitted to seven data points, $\nu=5$ and the acceptance threshold is
$\chi^2=13.08$. A fit with a $\chi^2$ value greater than this is rejected (with 99\% confidence).  The resulting
fits are shown in Figure \ref{fig-fits}, and the fitted parameter values---the spectral index $\alpha$ and the
reduced chi-squared value $\chi^2/\nu$---are shown in Table~\ref{tab-fit_results}.

\begin{table}
\caption{Results from power law fits to the extinction-corrected
  data. An asterisk indicates a fit rejected at the 99\% confidence
  level. Also shown are polarisation measurements from Impey \& Tapia
  (1990). Both their measurements, and the maximum polarisation that
  they found in the literature, are shown.}
\begin{tabular}{lcc|cc}
\hline
Source          & $\alpha$ &$\chi^2/\nu$  &$p$ (\%) &$p_{\rm max}$ (\%)\\
\hline
\hline
PKS\,0208$-$512  &$-0.82$   &$0.27$        &11.5     &11.5\\
PKS\,0506$-$612  &$-2.05$   &$11.21$*      &1.1      &1.1\\
PKS\,0528$+$134  &$-2.13$   &$2.52$        &0.3      &0.3\\
PKS\,1622$-$297  &$-1.90$   &$0.55$        &---      &---\\
PKS\,2230$+$114  &$-1.36$   &$0.74$        &7.3      &10.9\\
PKS\,2251$+$158  &$-1.12$   &$1.15$        &2.9      &16.0\\
\hline
\end{tabular}
\label{tab-fit_results}
\end{table}

We find that the fitted power laws can be separated into two groups. Three sources---PKS\,0506$-$612,
PKS\,0528$+$134, and PKS\,1622$-$297---are quite blue (although PKS\,0506$-$612 is a poor fit, as discussed below,
and should probably not be included in this group). These slopes are the same as those found at the blue end of
the distribution of fitted slopes for the Parkes Half-Jansky Flat-spectrum Sample (PHFS) quasars in Whiting,
Webster, \& Francis (2001). A natural interpretation of such blue slopes is that they are due to emission from the
accretion disc. This would be expected for high-redshift sources such as PKS\,0528$+$134, as the observed optical
emission is probing the rest-frame UV.

The remaining three sources have much redder optical slopes. Again
comparing to the PHFS sources in Whiting \etal (2001), we find that
these slopes are in the middle of the distribution, corresponding to
sources with synchrotron-dominated spectra.

One source had a poor fit: PKS\,0506$-$612. There is a large offset between the optical and NIR points, giving
what appears to be substantial spectral curvature. However, it is likely that the source has faded in the three
days between the two observations---if the spectrum is a smooth continuum then a reduction in flux of $\sim$ 1 mag
is implied. This also implies the intrinsic spectrum is a red power law of slope $\sim 1.3$. A similar, albeit
much weaker effect is seen in PKS\,0528$+$134, although this could also be due to the $I$ band data having excess
emission over the power law continuum. Contamination by the \mgii\ emission line (which would appear at
$\lambda_{\rm obs}=0.86\mu{\rm m}$) is a likely origin of this excess.

We can use data from the literature to support the interpretation of the different optical slopes. Synchrotron
radiation is intrinsically highly polarised, and so a significant net polarisation from a source is a good
indication that the emission is dominated by synchrotron emission. Impey \& Tapia (1990) measured polarisation for
five of the sources (excluding PKS\,1622$-$297), and their values are also shown in Figure \ref{tab-fit_results}.
It can be seen that the three sources with red optical slopes are the sources that have been observed to exhibit
high optical polarisations, indicating a likely synchrotron origin for the emission.

\section{Discussion}
\label{disc}

\subsection{Demographics}

All the sources we have studied here are radio-loud, flat-spectrum quasars (FSRQs). We would like to know how
typical these objects are compared to other FSRQs, or whether they are exceptional in some way that would explain
their gamma-ray emission. We examine here two properties---the redshift and the radio luminosity---and compare it
to a sample of flat-spectrum radio-loud objects. For this comparison sample we use the PHFS (Drinkwater \etal
1997), a sample of southern, radio-bright, flat-spectrum objects. We also include in this discussion the COMPTEL
sources that we did not observe: PKS\,1222$+$216 ($z=0.435$), PKS\,1226$+$023 (3C 273, $z=0.158$), and
PKS\,1253$-$055 (3C 279, $z=0.536$). Note that the latter two sources are also in the PHFS.

\begin{figure}[h]
\epsfbox{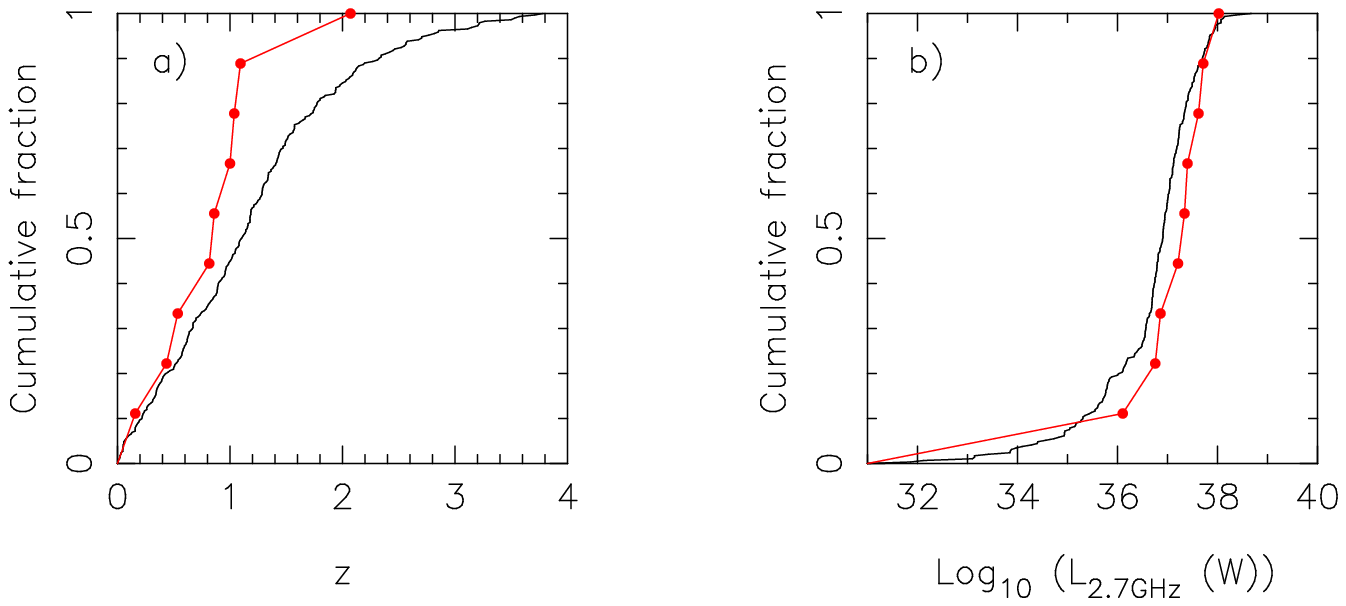}
\caption{ Comparison of the COMPTEL sample with the PHFS, with the
  black line representing the PHFS and the red line and points
  representing the COMPTEL sources. a) Cumulative redshift
  distributions.  b) Cumulative distributions of 2.7\,GHz radio
  luminosity. }
\label{fig-comp}
\end{figure}

The cumulative redshift distributions for both samples are plotted in Figure \ref{fig-comp}a. Note that we have
excluded from the PHFS distribution the 33 objects with no known redshift. These objects include a few BL Lac
objects, which have no emission line from which to measure a redshift, but are mostly optically faint, meaning no
suitable spectrum has been obtained. The COMPTEL sample tends to be biased towards lower redshift values, with
only one source at high redshift $z\gg 1$ (PKS\,0528$+$134 at $z=2.07$, Hunter \etal 1993). This is likely to be a
consequence of the sensitivity of the COMPTEL detector, with sources at suitably high redshift not being bright
enough at gamma rays for detection. We do note, however, that a Kolmogorov-Smirnov (K-S) test shows that the
distributions are different only at the 90\% confidence level (the probability that they are the same is 9.9\%).

Figure \ref{fig-comp}b shows the cumulative distributions of the radio luminosity for both samples. This is
calculated from the 2.7\,GHz fluxes from PKSCAT90 (assuming $H_0=75$\,km\,s$^{-1}$\,Mpc$^{-1}$ and $q_0=0.5$). We
see that the COMPTEL sources tend to have, on average, higher radio luminosities than the PHFS sources (there is a
difference in average luminosity of 0.56 dex, or a factor of $\sim3.6$). Again, however, a K-S test shows this is
not very significant (the probability that the distributions are different is 12.6\%). We can, however, discuss
the trends shown in the graphs.

Even though, as we see from the redshift distribution, the COMPTEL sample lacks the high-redshift sources (which
one would expect to be more luminous than low-redshift sources in a flux-limited sample), the average radio
luminosity of this sample is still greater than that of the PHFS. This implies that the sample selected at gamma
rays lacks sources of low radio luminosity. This is probably an indication that a certain radio luminosity is
required for a source to be active at high-energy gamma rays.

\subsection{Multiwavelength SEDs}
\begin{figure}[h]
\epsfbox{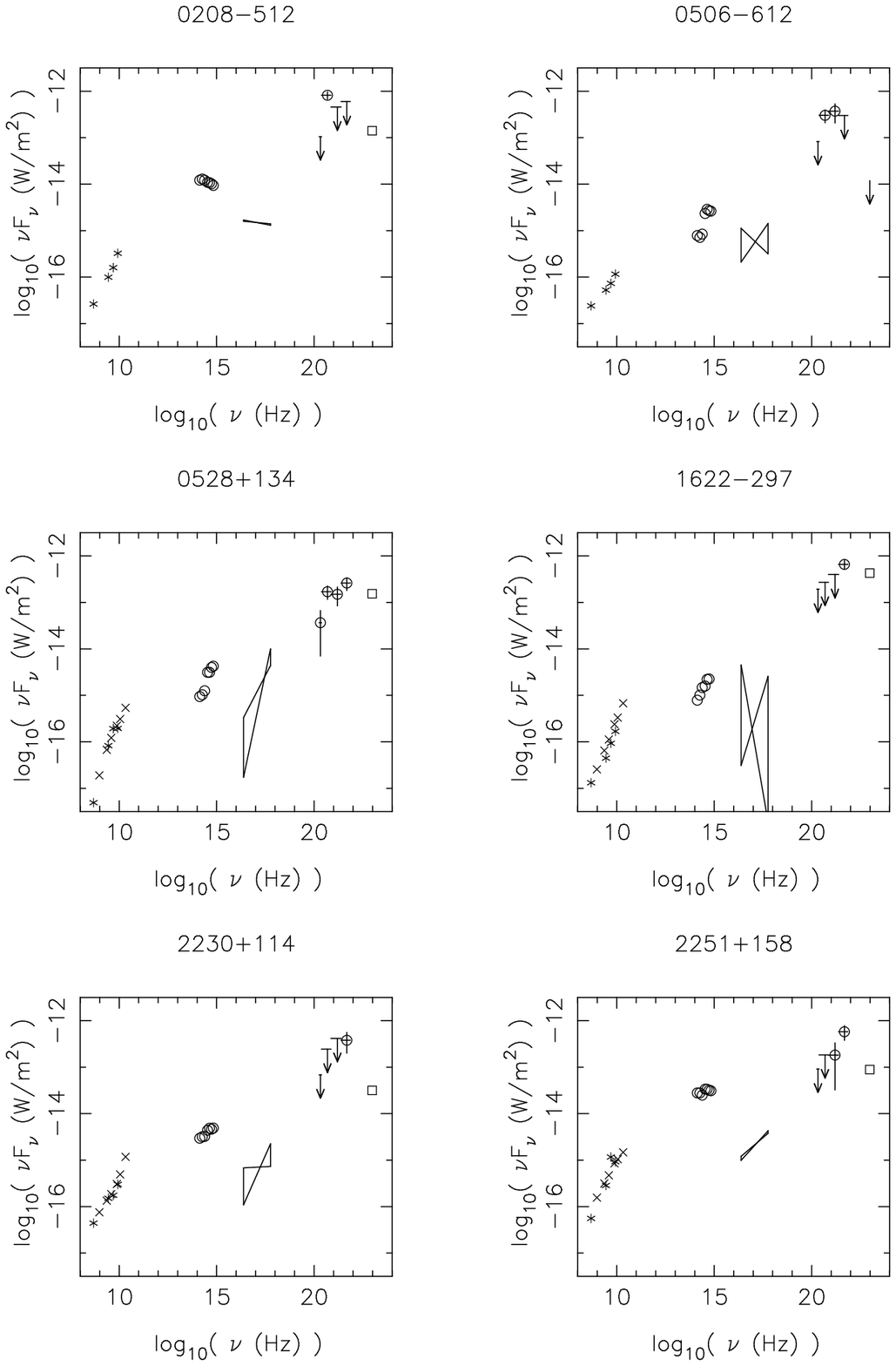}
\caption{ Broad-band SEDs for the six sources discussed in this
  paper. See text for references for each dataset. Upper limits (\ie
  non-detections) for the high-energy data are indicated by
  arrows. Note that the different sets of observations are in general
  not simultaneous.}
\label{fig-seds}
\end{figure}

In Figure \ref{fig-seds} we plot the multiwavelength (radio to gamma-ray) SEDs constructed from published data, as
well as the new data presented herein. Two epochs of radio observation are shown. One set of data (indicated by
the asterisks) is from PKSCAT90, while the second (indicated by crosses) is from Kovalev \etal (1999). The latter
does not include PKS\,0208$-$512 and PKS\,0506$-$612. The optical/NIR data (from this paper) have been corrected
for Galactic extinction. The X-ray data come from Brinkmann, Yuan, \& Siebert (1997). The `bowtie' represents the
twin constraints given by the errors on the total 0.1--2.4\,keV flux and the spectral slope. (Note that the
constraints on the slope for PKS\,1622$-$297 are rather poor.) The COMPTEL data, taken from Sch{\"o}nfelder \etal
(2000), are indicated by the four points between $10^{20}$\,Hz and $10^{22}$\,Hz. The horizontal error bars
indicate the energy range of each bin, while the vertical error bars indicate the flux error. Upper limits
($2\sigma$) are indicated by arrows. The final point is the EGRET datum from the Third EGRET Catalog (Hartman
\etal 1999), using their approximation to generate a 400\,MeV flux density by multiplying the catalogued flux by
1.7. Note that the observations presented here are not simultaneous.

The SEDs provide an interesting glimpse of the range of different spectral shapes present in radio-loud quasars.
The optically-red sources, particularly PKS\,0208$-$512 and PKS\,2251$+$158, show optical--NIR SEDs that can quite
easily be connected to the radio data, as would be expected if both regimes are dominated by synchrotron emission
from the relativistic jet.

For other sources, however, it is less obvious what, if any, connection exists between these two regimes. For
instance, the $K$ band flux of PKS\,1622$-$297 is approximately equal (on a $\nu F_\nu$ scale) to the 20\,GHz flux
from Kovalev \etal (1999), while the spectral index in the two regimes is approximately the same. This, however,
should not be surprising if we attribute the optical emission to the accretion disc, as it will then not be the
same component as the radio emission. In such a picture, the synchrotron emission from the jet has turned over
somewhere in the infrared, leaving the accretion disc emission to dominate the optical. Note that the flux level
at radio frequencies is approximately equal for all sources, whereas the redder sources tend to be brighter in the
optical/NIR, indicating the presence of an additional emission component.

Clearly, while the non-thermal processes (\ie synchrotron and IC) are undoubtedly important for the radio and
high-energy emission, the optical regime is not always going to be dominated by them. This was shown in the
detailed study of the optical/NIR properties of radio-loud quasars by Whiting \etal (2001), which found that a
large fraction of sources showed no evidence for non-thermal synchrotron emission at optical wavelengths. The
observations of strong emission-line spectra and low polarisation in the optical for the blue objects in this
sample also back this up.

This raises an important issue when it comes to modelling the multiwavelength SEDs of such objects. Detailed
models do exist in the literature, and are used with good effect to model the high-energy emission (X-rays through
gamma rays). See, for instance, Mukherjee \etal (1999) for modelling of the multiwavelength spectra of
PKS\,0528$+$134. However, many of these models do not take the lower-energy (\ie radio through optical) data fully
into account. As an example, we refer the reader to Figures 8 and 9 of Mukherjee \etal (1999), where the model
flux in the optical goes in the opposite direction to the data. This seems to be due to the fitted models not
fully taking into account the presence of significant emission from the accretion disc.

This is an important issue, since if the thermal accretion disc emission is dominating the optical, the
synchrotron emission in that frequency range is much less than previously assumed. This will then affect the
derived values for the Lorentz factors, which affect the generation of the gamma rays via the IC process.

The photon density at the jet of the seed photons is another key
parameter for calculating the IC luminosity. For a given
observed total optical flux, this parameter will be different for
synchrotron-dominated and thermally-dominated spectra, and its
spectral shape will differ as well. This will alter the predicted
shape of the IC emission.

It can be seen, therefore, that a good understanding of the makeup of the optical emission is crucial for accurate
modelling of the high-energy emission. This is something that needs to be considered for future multiwavelength
campaigns involving high-energy observations and subsequent modelling, particularly with the advent of new
gamma-ray telescopes such as INTEGRAL and GLAST.

\section{Conclusions}

We have observed six COMPTEL-detected quasars at optical and NIR wavelengths, and we present photometry in seven
wavebands, corrected for Galactic extinction.

We find large differences in the optical properties between the sources. Two of the sources have intrinsically
blue optical SEDs, and are likely to be dominated by thermal emission from an accretion disc. A further three have
much redder optical slopes, and are more likely to be dominated in the optical by non-thermal synchrotron emission
from the relativistic jet.

This identification of accretion disc emission in some gamma-ray sources has important implications for the
sources' broad-band modelling. It implies both that the thermal emission is stronger and the non-thermal emission
weaker in the optical regime than previously considered, indicating that broad-band models used to explain the
high-energy emission will need to be revisited, particularly as far as the optical regime is concerned.

\section*{Acknowledgments}

The authors would like to thank A.\ Melatos for some very helpful discussions.  We also want to thank the MSSSO
TAC for granting us time for the observations at the 40\,in and 2.3\,m telescopes at the SSO.

This research has made use of the NASA/IPAC Extragalactic Database
(NED) which is operated by the Jet Propulsion Laboratory, California
Institute of Technology, under contract with the National Aeronautics
and Space Administration.

\newpage
\section*{References}

% PASA uses the same conventions as ApJ for journal abbreviations.  Sample
% references are as follows.
% Please follow the same format for your references.

%\reference Author, A. B., Anotherauthor, C. D. \and Thirdauthor, E. F. 1990,
% PASA, 7, 350

% for a journal article, or

% \reference Author, A.B. \and Anotherauthor, C. D. 1990, in This is a
% Book %Title, ed. C. D. Editor, (City: Publisher Name), 437

% for a book.

\reference Bersanelli, M., Bouchet, P., \& Falomo, R. 1991, A\&A, 252, 854

\reference Bessell, M.S., Castelli, F., \& Plez, B. 1998, A\&A, 333, 231

\reference Brinkmann, W., Yuan, W., \& Siebert, J. 1997, A\&A, 319, 413

\reference Bloom, S.D., \& Marscher, A.P. 1996, ApJ, 461, 657

\reference Carter, B.S., \& Meadows, V.S. 1995, MNRAS, 276, 734

\reference Collmar, W. 2001, in High Energy Gamma-Ray Astronomy, AIP Proceedings Volume 558, ed.\ F.A.\ Aharonian
$\&$ H.J.\ V{\"o}lk (Melville, NY: American Institute of Physics), 656

\reference Drinkwater, M.J., \etal 1997, MNRAS, 284, 85

\reference Graham, J.A. 1982, PASP, 94, 244

\reference Hartman, R.C., \etal 1999, ApJS, 123, 79

\reference Hunter, S.D., \etal 1993, ApJ, 409, 134

\reference Impey, C.D., \& Tapia, S. 1990, ApJ, 354, 124

\reference Kovalev, Y.Y., Nizhelsky, N.A., Kovalev, Yu.A., Berlin, A.B., Zhekanis, G.V., Mingaliev, M.G., \&
Bogdantsov, A.V.\ 1999, A\&AS, 139, 545

\reference McGregor, P., \etal 1994, in Infrared Astronomy with Arrays: The Next Generation, ed.\ I.S. McLean
(Dordrecht: Kluwer), 299

\reference Mukherjee, R., \etal 1999, ApJ, 527, 132

\reference Raiteri, C.M., Villata, M., Lanteri, L., Cavallone, M., \& Sobrito, G.\ 1998, A\&AS, 130, 495

\reference Schlegel, D.J., Finkbeiner, D.P., \& Davis, M. 1998, ApJ, 500, 525

\reference Sch{\"o}nfelder, V., \etal 2000, A\&AS, 143, 145

\reference Whiting, M.T., Webster, R.L., \& Francis, P.J. 2001, MNRAS, 323, 718

\reference Wright, A., \& Otrupcek, R.\ 1990, Parkes Catalogue (Australian Telescope National Facility) (PKSCAT90)

\reference Zhang, Y.F., Marscher, A.P., Aller, H.D., Aller, M.F., Terasranta, H., \& Valtaoja, E.\ 1994, ApJ, 432,
91

\end{document}